# Polarized indistinguishable single photons from a quantum dot in an elliptical micropillar

Yu-Ming He[1,2,*], Hui Wang[1,2,*], Stefan Gerhardt[3,*], Karol Winkler[3], Jonathan Jurkat[3], Ying Yu[4], Ming-Cheng Chen[1,2], Xing Ding[1,2], Si Chen[1,2], Jin Qian[1,2], Zhao-Chen Duan[1,2], Jin-Peng Li[1,2], Lin-Jun Wang[1], Yong-Heng Huo[1,2], Siyuan Yu[4], Sven Höfling[3,5], Chao-Yang Lu[1,2], Jian-Wei Pan[1,2]

* These authors contributed equally to the work

[1] *Shanghai branch, National Laboratory for Physical Sciences at Microscale and Department of Modern Physics, University of Science and Technology of China, Shanghai 201315, China*

[2] *CAS Centre for Excellence and Synergetic Innovation Centre in Quantum Information and Quantum Physics, University of Science and Technology of China, Hefei, Anhui 230026, China*

[3] *Technische Physik, Physikalisches Instität and Wilhelm Conrad Röntgen-Center for Complex Material Systems, Universitat Würzburg, Am Hubland, D-97074 Würzburg, Germany*

[4] *State Key Laboratory of Optoelectronic Materials and Technologies, School of Electronics and Information Technology, School of Physics, Sun Yat-sen University, Guangzhou 510275, China*

[5] *SUPA, School of Physics and Astronomy, University of St. Andrews, St. Andrews KY16 9SS, United Kingdom*

**The key challenge to scalable optical quantum computing[1], boson sampling[2], and quantum metrology[3] is sources of single photons with near-unity system efficiency and simultaneously near-perfect indistinguishability in all degrees of freedom[4] (including spectral, temporal, spatial, and polarization). However, previous high-indistinguishability solid-state single-photon sources[4-6] had to rely on polarization filtering that reduced the system efficiency by at least 50%. Here, we overcome this challenge by developing a new single-photon source based on a coherently driven quantum dot embedded in an elliptical micropillar. The asymmetric cavity lifts the polarization degeneracy into two orthogonal linearly polarized modes with a suitable energy separation[7]. We design an excitation-collection scheme that allows the creation and collection of single photons with an indistinguishability of 0.976(1) and a degree of polarization of 91%. Our method provides a solution of combining near-unity system efficiency and indistinguishability compatible with background-**

**free resonant excitation, and opens the way to truly optimal single-photon sources for scalable photonic quantum technologies.**

Single photons are appealing candidates for quantum communications[8], quantum-enhanced metrology[3] and quantum computing[1,2]. In view of the quantum information applications, the single photons are required to be controllably prepared with a high efficiency into a given quantum state. Specifically, the single photons should possess the same polarization, spatial mode, and transform-limited spectro-temporal profile for a high-visibility Hong-Ou-Mandel-type quantum interference[8,9]. However, compatibly achieving near-unity system efficiency and photon indistinguishability remained an important challenge.

Self-assembled quantum dots show so far the highest quantum efficiency among solid-state quantum emitters and thus can potentially serve as an ideal single-photon source[4,10-14]. There has been encouraging progress in recent years in developing high-performance single-photon sources[4]. Pulsed resonant excitation on single quantum dots has been developed to eliminate dephasing and time jitter, which created single photons with near-unity indistinguishability[15]. Further, by combining the resonant excitation with Purcell-enhanced micropillars[5,6] or photonic crystals[16], the generated transform-limited[17,18] single photons have been efficiently extracted out of the bulk GaAs (with an extraction efficiency of up to 82% in ref. 19) and funneled into a single mode at far field. Although such single-photon sources have been applied in multi-photon boson sampling experiments and shown a significant advantage[19] compared to parametric down-conversion sources, there remained a critical drawback that prevented a scalable implementation.

The main overhead of the resonant excitation is suppression of the scattering from the excitation laser which has the same wavelength of the quantum dot single photons. So far, the most effective method is filtering in the polarization degree of freedom[15,20], where a linearly polarized laser excites a single-electron charged quantum dot, and the collection arm, another polarizer with its polarization perpendicular to the excitation light extinguishes the scattered laser background. However, such a cross-polarization method meanwhile reduced the system efficiency of the single photons which were right and left circularly polarized by at least 50%.

One possible remedy is using orthogonal excitation and collection, that is, exciting from the side into the waveguide mode of the micropillar cavity and collecting from the top. While such a method had been implemented in earlier experiments[21,22], it didn't allow the realizations of high-performance single-photon sources yet. More importantly, in the previous experiments[5,6] that relied on doubly degenerate excitonic transitions in singly-charged quantum dots, the emitted single photons are randomly polarized at right or left circular. Polarization filtering can be used to project the quantum states at a fixed polarization, which again sacrifices the efficiency by 50% and thus intrinsically limits scalable applications in optical quantum computing[1], boson sampling[2] and quantum metrology[3].

It is thus crucial to develop a polarized single-photon source with simultaneously high system efficiency and near-unity indistinguishability. To do so, we need to break the original polarization symmetry of the quantum dot emission, and develop a new way for laser-background-free pulsed resonance fluorescence without sacrificing the system efficiency. Here, we design a feasible proposal satisfying these conditions, and we report its experimental demonstration.

A general framework of our protocol is shown in Fig. 1a. The key idea is to couple a single quantum dot to a geometrically birefringent cavity in the Purcell regime. The asymmetric microcavity is designed such that it lifts the polarization degeneracy of the fundamental mode and splits it into two orthogonal linearly polarized—horizontal ($H$) and vertical ($V$)—modes, with a cavity linewidth of $\delta\omega$ and a frequency separation of $\Delta\omega$. Suppose a single-electron charged quantum dot, which is a degenerate two-level system, is brought into resonance with the cavity $H$ mode, and off-resonant with the cavity $V$ mode with a detuning of $\Delta\omega$. The cavity redistributes the spontaneous emission rate of the quantum emitter into the $H$ and $V$ polarizations with a ratio of $1+4(\Delta\omega/\delta\omega)^2:1$. For a series of realistic Purcell factors, the extraction efficiency of polarized single photons is plotted in Fig. 1b. For example, with a Purcell factor of 20 and $\Delta\omega/\delta\omega=3$, the polarized single-photon efficiency is 93%.

To ensure a high photon indistinguishability, it is essential to resonantly drive the quantum dot transition. The excitation laser's polarization is assigned to be *V*, while in the output, an *H* polarizer—aligned with the dominant polarization of the cavity-enhanced single photons—is used to extinguish the excitation laser scattering. Thus, our protocol is compatible with the cross-polarization technique and has little loss of single-photon system efficiency. We note that, however, our protocol does come with a small price. As the excitation laser resonant at the emitter's transition is off-resonantly coupled to the *V* polarization cavity mode (see Fig. 1a), thus its driving power needs to be stronger for a deterministic π pulse. Compared to the case using an isotropic microcavity, the increase factor is calculated and plotted in Fig. 1b. For example, at $\Delta\omega/\delta\omega = 3$, the excitation laser power is estimated to be ~10 times higher.

Our protocol is applicable in various photonic structures[10,11] such as micropillars[5,6] and photonic crystals[16]. Here, for the first time, we demonstrate a coherently pumped single quantum dot embedded in an elliptical micropillar as a high-efficiency single-photon source with near-unity polarization and indistinguishability. The GaAs/InAs micropillars with elliptical cross section were first studied by Gayral *et al.* in 1998, who observed splitting of the degenerate fundamental mode of the cavity[7]. It was shown that the two split modes were linearly polarized, aligned in parallel with the major and minor axis of the elliptical cross section[23,24]. Single GaAs/InAs quantum dots embedded in such elliptical micropillars showed polarization-dependent Purcell enhancement[25,26] and single photons could be preferentially generated in a single polarization state[23-27]. However, all the previous experiments with elliptical micropillars were performed with non-resonant laser excitation, which degraded the purity of the created single photons. Moreover, neither high collection efficiency nor high photon indistinguishability was realized in the previous elliptical micropillar devices.

In our experiment, we use a single InAs/GaAs quantum dot embedded in a λ-thick micropillar with an elliptical cross section (see Fig. 2a), which has a major (minor) axis of 2.1 µm (1.4 µm). The quantum dot is sandwiched between 25.5 (15) λ/4-thick AlAs/GaAs mirror pairs forming the lower (upper) distributed Bragg reflectors. The sample is placed inside a bath cryostat with a lowest temperature of 1.5 K. A confocal microscopy is used to excite the quantum dot and collect the emitted single photons.

We first characterize the elliptical micropillar using non-resonant excitation with a ~780 nm c.w. laser at high power. Figure 2b shows two nondegenerate fundamental cavity modes[7,24-27], at 873.69 nm (labelled as $M_1$) and 874.13 nm (labelled as $M_2$), respectively, with a splitting of 171 GHz. The $M_1$ and $M_2$ modes correspond to the minor and major axis, with quality factors of 4621 and 5521, respectively. A modest reduction of the quality factor of $M_1$ compared to $M_2$ is due to the smaller micropillar diameter. Polarization-resolved measurements (see Fig. 2c) confirms the polarization of $M_1$ ($M_2$) is parallel to the minor (major) axis which we label as *V* (*H*), with high degrees of polarization of 94.5% (99.1%). This symmetry-broken, highly-polarized microcavity is the key to generate pure, single-polarized single photons in the present work.

Another important issue is that the emitted photons should be efficiently collected into a single-mode fiber. It is known that Purcell-enhanced micropillars with circular cross sections are favorable for directional, nearly Gaussian far field[4-6]. We simulate the far field of our elliptical micropillar using finite-different time-domain method. The numerical results in Fig. 2d indicates that the far-field distribution of fundamental mode inherits the cross-section pattern of the elliptical pillar. For the parameters used in our micropillar (1.4-2.1 µm), the overlap between the simulated distribution and a perfect Gaussian is ~98%. This indicates that a suitable ellipticity only slightly affects the collection efficiency. In addition, as shown in Fig. 2d, the single-photon emission is highly directional. An objective lens with a numerical aperture of 0.65 is capable of collecting ~99% of the emitted photons.

By temperature tuning, we bring the singly-charged quantum dot into resonance with the $M_2$ cavity mode (see Fig. 2b) at ~4 K. A pulsed laser with a central wavelength of ~874.13 nm and pulse duration of ~3 ps resonantly excites the quantum dot embedded in the micropillar. Following our protocol shown in Fig. 1a, driven by a *V* polarized laser, $M_2$ cavity Purcell-enhanced *H* polarized single photons are collected in the output of the microscope with a ~$10^7$:1 cross-polarization extinction of the scattering laser background. Under a driving laser power of ~5 nW at π pulse and a repetition rate of

76.1 MHz, about 7.8 million resonance fluorescence single photons per second are detected by a single-mode fiber-coupled superconducting nanowire single-photon detector with an efficiency of ~70% (see Fig. 3a for a full Rabi oscillation curve). As a comparison experiment, we excite the quantum dot with *H* polarized laser and collect *V* polarized single photons. As shown in Fig. 3b, at π pulse, only about 0.36 million single photons per second are detected. Therefore, we estimate a degree of polarization —defined as $(I_H - I_V)/(I_H + I_V)$, where $I_H$ ($I_V$) denotes the detected intensity of *H* (*V*) polarized photons—of 91.2% of the generated single photons from this elliptical cavity device. For quantum information applications requiring a pure polarization state, such a single-photon source will induce a loss of only 4.4% due to polarization, thus overcoming the fundamental 50% loss using the previous circular micropillars[5,6]. As a price, we note that the driving power at π pulse in Fig. 3a, due to off-resonant cavity coupling, is about 10 times higher than that in Fig. 3b where the driving laser is resonant to the cavity, which is in good agreement with the theoretical prediction in Fig. 1b.

Finally, we experimentally test the purity and the indistinguishability of the generated single photons. Measured with a standard Hanbury Brown and Twiss setup, the second-order correlation is displayed in Fig. 4a which shows $g^2(0) = 0.050(2)$ at zero time delay. The imperfection of the measured $g^2(0)$ is mainly due to laser leakage. The photon indistinguishability is characterized by a Hong-Ou-Mandel interferometer with the time separation between the two consecutive laser pulses—and thus the two emitted single photons—set at 13 ns. Figure 4b shows the photon correlation histograms of normalized two-photon counts for orthogonal and parallel polarizations. The observed contrast of the counts for the two cases at zero delay can be used to extract a raw two-photon interference visibility of 0.911(1). Taking into account of the imperfect single-photon purity and the unbalanced (47:53) beam splitting ratio in the optical setup, we can calculate a corrected photon indistinguishability of 0.976(1). These results indicate that the near-unity single-photon generation efficiency, purity, indistinguishability can be simultaneously combined with high degrees of polarization in a coherently driven single-photon device.

For the end users in quantum information, it is the single-photon system efficiency that ultimately matters. For instance, a system efficiency above 50% is necessary for the boson sampling[2] to show quantum advantages over classical computers. While our current single-photon source has overcome the 50% loss due to polarization filtering, its final system efficiency is about 15%, which is mainly limited by the Purcell factor of 3.4, scattering loss (~20%) due to imperfect micropillar sidewall, single-mode fiber coupling loss (~30%), and transmission loss (~47%) in the optical setup. Future work can employ far-field lithography[28,29] or imaging[30] technique to deterministically position the quantum dot at the center of the micropillars, which can yield a larger Purcell factor and more directional emission. Our protocol can also be adopted to engineer asymmetric photonic crystals[16] and Bragg gratings[31] which can have much small cavity volumes than the typical micropillars and thus higher Purcell factors.

**Figure captions:**

**Figure 1: The theoretical scheme of a polarized single-photon source by resonantly pumping a quantum emitter in a birefringent microcavity**. **a.** The asymmetric cavity supports two-fold non-degenerate cavity modes, one in horizontal (*H*) polarization and the other one in vertical (*V*) polarization. We bring the emitter into resonance to the *H* mode, at which single photons are preferentially prepared due to polarization-selective Purcell enhancement. The *V* cavity mode, suitably detuned from the emitter, serves for laser excitation. A *V* polarized laser pulse is weakly coupled to the *H* mode due to a small overlap between the two cavity modes, thus can pump the two-level system to its excited state with a higher power. **b.** The system efficiency of preparing single polarized single photons as a function of the ratio of the cavity splitting to the cavity linewidth. Examples with four different Purcell factors are plotted. The dashed line shows the increased factor of the pump laser power required for a deterministic $\pi$ pulse, compared to the case of micropillar with circular cross section.

**Figure 2: Characterization of the elliptical micropillar**. **a**. An illustration of InGaAs quantum dot-elliptical micropillar device used in this work, which has a major (minor) axis of 2.1 µm (1.4 µm). **b**. Two fundamental mode of the elliptical micropillar, $M_1$ and $M_2$, with a splitting of 171 GHz. The linewidths of the $M_1$ and $M_2$ are 74 GHz and 62 GHz, respectively. The quantum dot is resonant with the $M_2$ at a temperature of 4 K. **c**. Polarization-resolved measurement of the two cavity modes, which are perpendicular with each other. The degree of polarization of the $M_1$ and $M_2$ modes is 94.5% and 99.1%, respectively. **d**. Numerical simulation of far-field distribution of the electrical intensity of the emission, which inherits the cross section of elliptical micropillars.

**Figure 3: The deterministic generation of polarized single photons under resonant excitation.** **a**, The polarization of the excitation laser is set to be parallel to the minor axis, and the output polarizer is aligned parallel to the major axis, that is, orthogonal to the input laser polarization in order to extinguish the laser background. The measured

pulsed resonance fluorescence single photon counts are plotted as a function of the laser power, which shows a clear Rabi oscillation. Under π pulse at ~5 nW, we observe ~7.8 million single-photon counts per second by a superconducting nanowire single-photon detector. **b,** For a controlled experiment, we exchange the polarizations of the excitation laser and the collected single photons. In this case, the single photons are suppressed by the cavity. Under π pulse at ~0.51 nW, only 0.36 million single photons per second are detected.

**Figure 4: The single-photon purity and indistinguishability. a.** Measurement of the second-order correlation function, which gives $g^2(0) = 0.050(2)$. **b,** Characterization of photon indistinguishability in a Hong-Ou-Mandel interferometer with a time delay of 13 ns. A significant count suppression at zero delay is observed when the two photons are in parallel polarization (red dots) compared to the case when the two photons are in cross polarization (blue circles). After correction, the calculated indistinguishability is 97.6%.

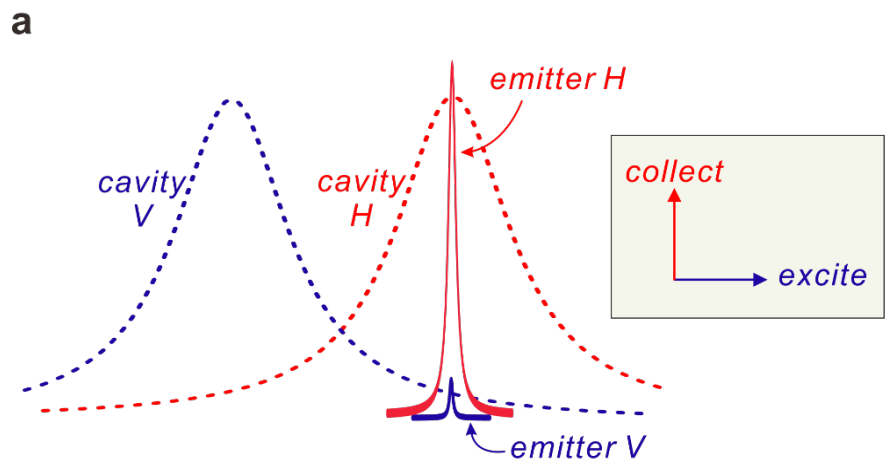 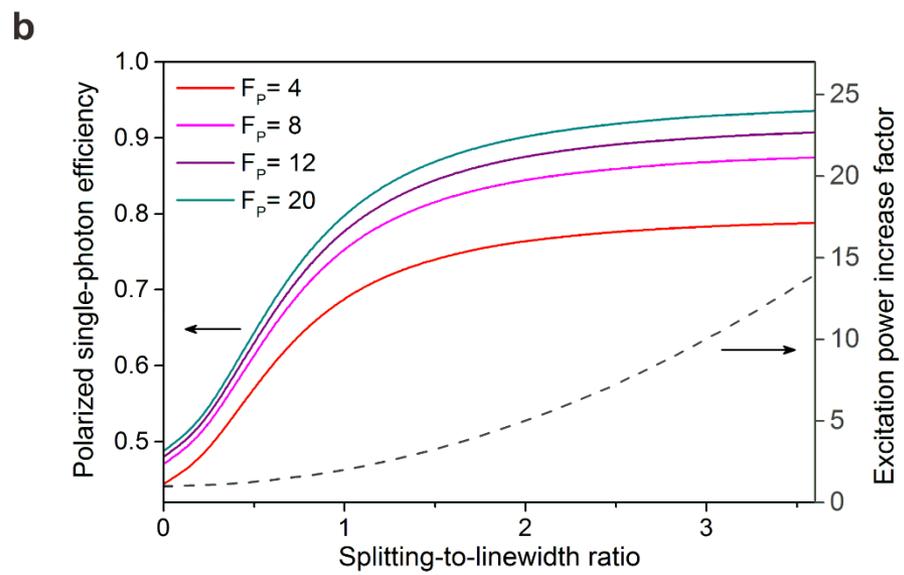

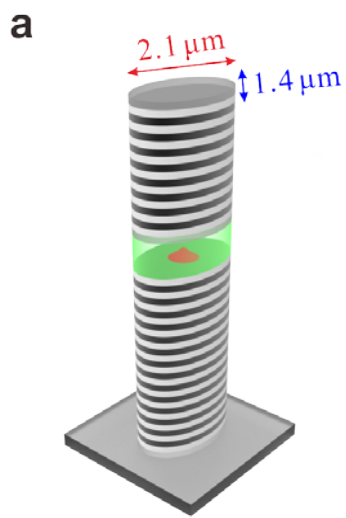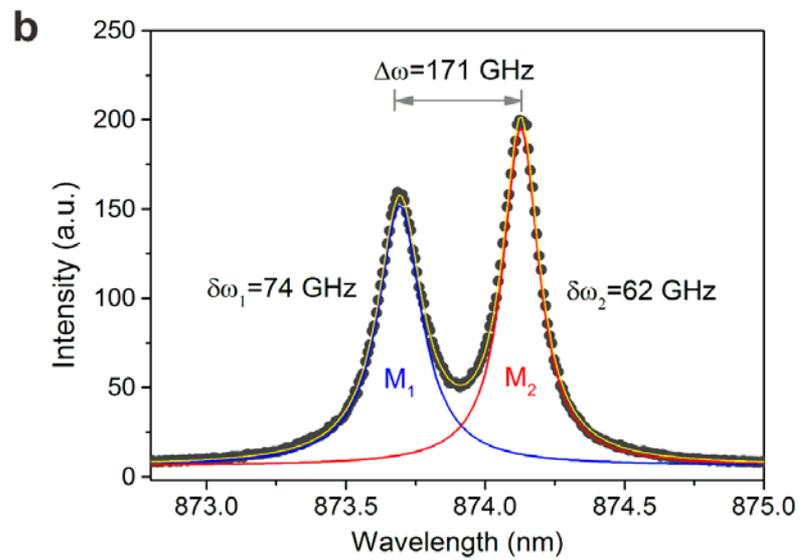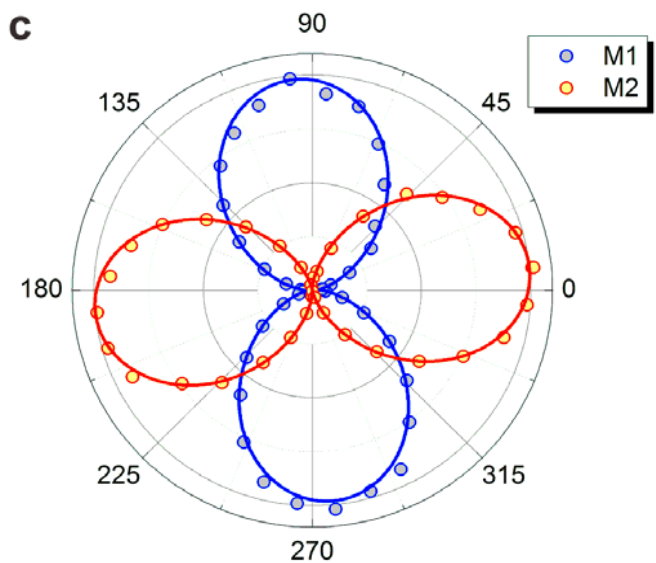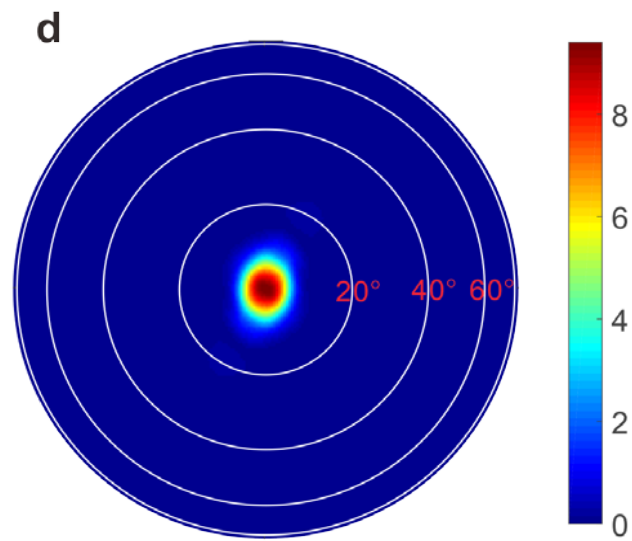

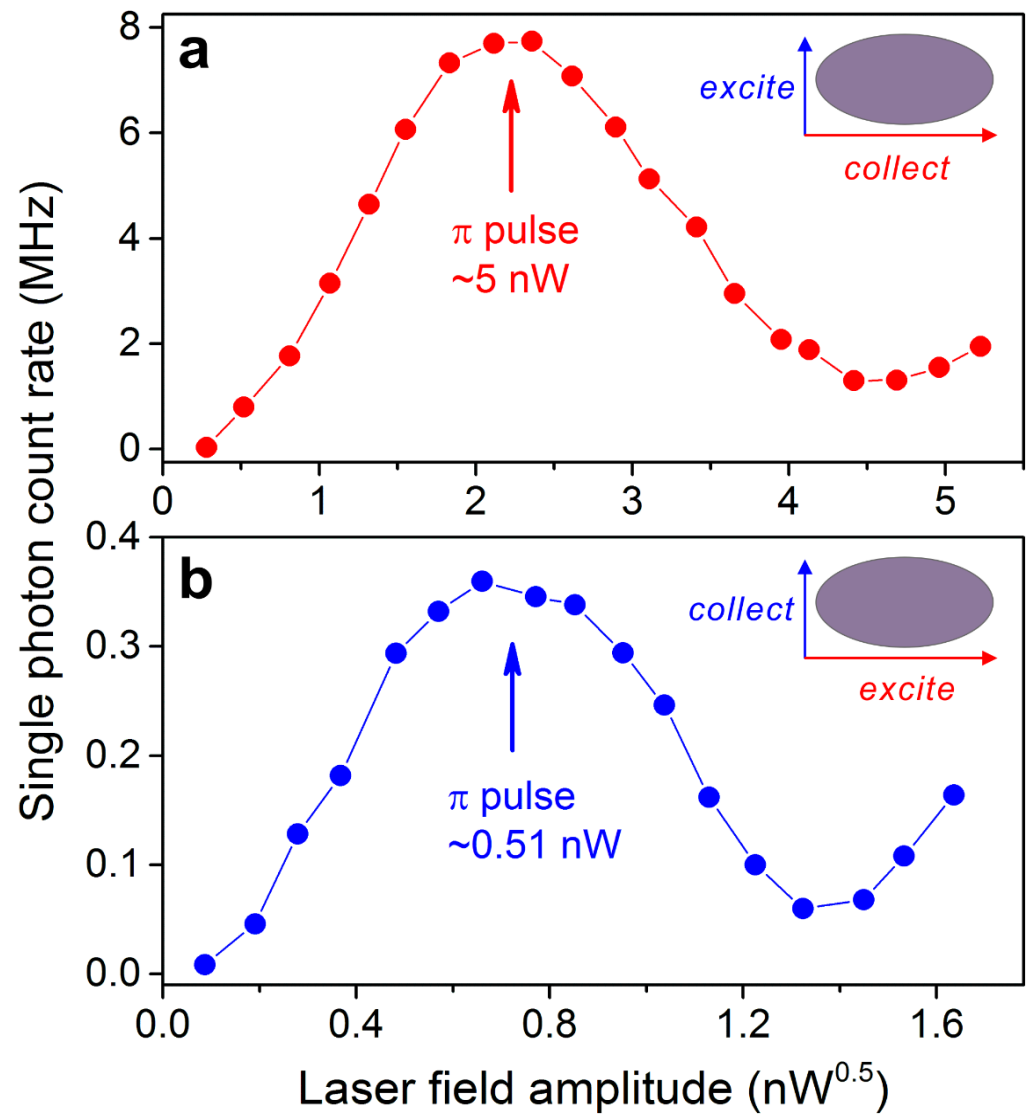

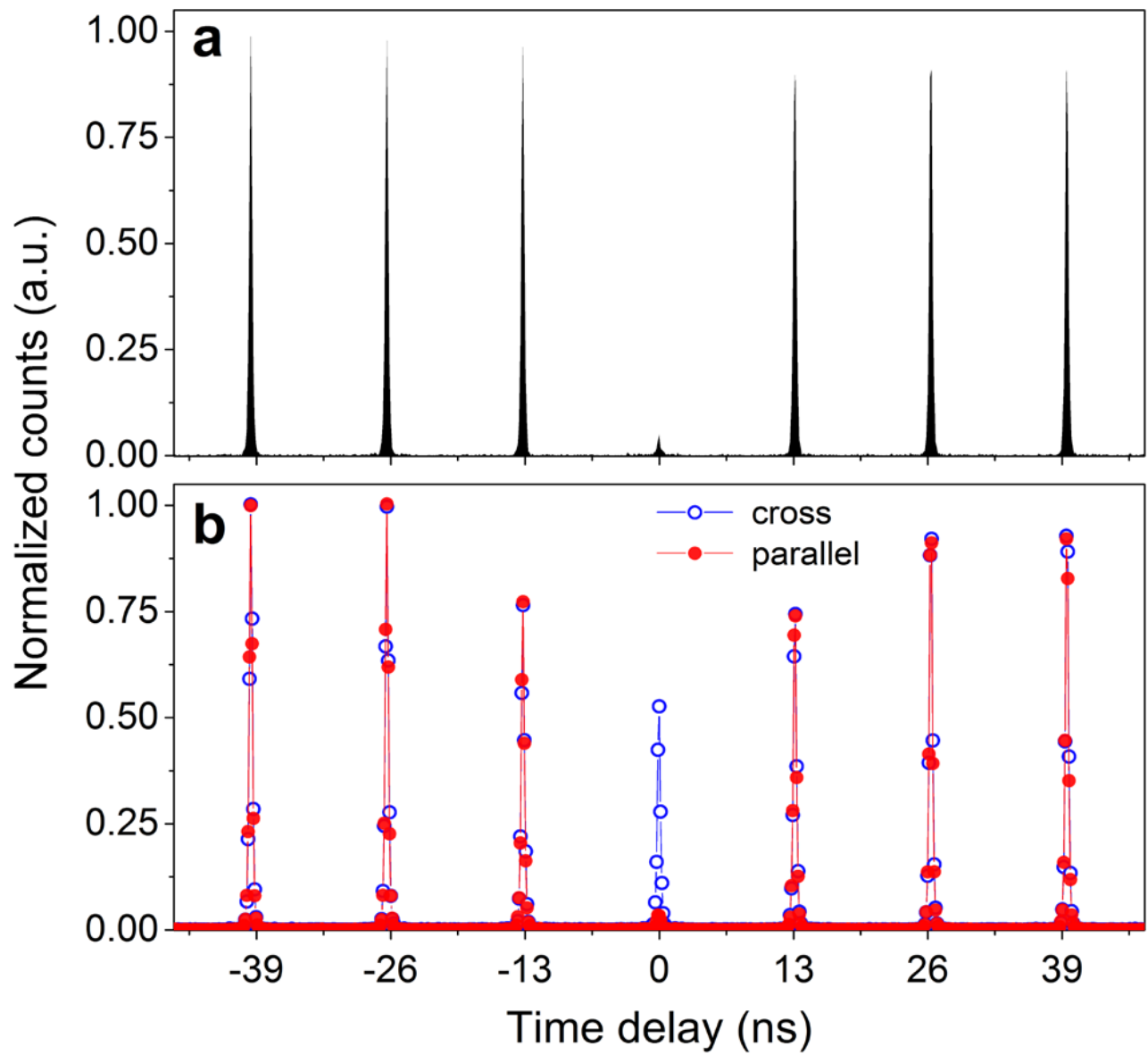